# Roadmap for ILC Detector R&D Test Beams

*World Wide ILC Detector R&D Community*

October 17, 2007

**Abstract**

This document provides a roadmap for ILC Detector test beam needs in the next 3 – 5 years. In this period, detector Letters of Intent are expected by fall 2008, the ILC Engineering Design Report to be submitted in early 2010 (with detector Technical Design Reports soon thereafter) and funding approval to construct the ILC and its detectors in 2012. ILC Detectors are required to have unprecedented precision to be able to elucidate new physics discoveries at TeV energies from the LHC and ILC machines, and to fully exploit experimental investigation at the electroweak unification energy scale. Achieving this requires significant investment for detector test beam activities to complete the R&D needed, to test prototypes and (later) to qualify final detector system designs, including integrated system tests. This roadmap document describes the need for a significant increase in resources for ILC test beam activities. It should be used by test beam facility managers and the worldwide ILC leadership to assure that the necessary resources and facilities are made available to meet the needs in time.

# Executive Summary

This document presents an ILC detector test beam roadmap. The information contained in this document is primarily a result of the ILC test beam workshop held at Fermilab in January 2007 [1] and the information collected as recently as LCWS07 at DESY. Given the nature of rapidly progressing detector R&D activities and the corresponding needs, this document is expected to be updated regularly as the needs arise. In this section, however, we provide an executive summary of this roadmap document and some explicit recommendations.

- Since it is ideal to keep the ILC detector timeline synchronous to that of the accelerator, the detector R&D time line is dictated by that spearheaded by the GDE in accelerators.
- The latest ILC detector roadmap presented, by the WWS and ILCSC at LCWS07 in DESY, re-emphasizes the importance of this synchronization and provides a path to two ILC detector collaborations for detector TDRs ready in time with the accelerator TDR.
- These two significant events in determining the ILC detector timeline naturally demand significant resources and support to expeditiously complete detector R&D and corresponding beam test activities for an informed decision making. Thus the demand on detector beam test facilities will grow significantly.
- A large number of tests for detector characterization can be carried out with low energy beam particles, such as electrons and hadrons, of moderate rates.
- For large scale, reliable performance tests, however, the range in required beam particle species and momentum, along with the data collection rates, grow significantly.
- The number of detector R&D groups grew significantly during the past year or so due primarily to the increase in the number of activities in tracking detectors and vertex sensor technology development groups. The number of groups requiring beam will grow further during the next 3 – 5 year period covered in this document.
- The demand in calorimeter groups kept increasing at a steady rate and their physics prototypes' presence in beam is becoming realized and is expected to grow moderately.
- The muon detector groups have been keeping up their beam test activities but are not expected to grow significantly given smaller number of groups working on this detector system. In addition, this detector group has higher probability of coexisting with other detector systems throughout the beam tests.
- The beam instrumentation and machine detector interface groups have been most active in carrying out beam tests at SLAC's ESA and KEK's ATF facilities. The level of activity is expected to be at about the same level or increase moderately through the next 3 – 5 year period.
- Several independent sub-detector R&D groups have joint forces to leverage scarce resources by developing common systems, most notably the DAQ, allowing groups to perform beam tests without reinventing wheels. An example is the CALICE collaboration where many different calorimeter technology groups utilize common DAQ hardware and software systems, online monitoring, mechanical support structure, reconstruction software and analysis software. This allows these groups to extract physics results from their detector technologies, critical for the ILC detector design, in an expedient manner.
- Despite the anticipated increase in ILC detector R&D test beam needs, the number of suitable facilities that can meet these needs will reduce over the next 3 – 5 year period.
  - SLAC ESA ILC test beam program is uncertain beyond FY08. SLAC's End Station A facility, with its unique capability for a high energy primary electron beam that matches many of the ILC beam parameters, is an important facility for BI and MDI R&D activities. Shutting down this facility after FY08 limits BI&MDI group activities to KEK's ATF which has to be shared with other accelerator R&D efforts.
  - CERN's PS and SPS can provide a variety of particle species with a wide momentum range. The CERN program foresees the continuation of the PS and SPS fixed target and test beam program during the LHC era, and measures are taken to deliver beams most of the time even during commissioning and operation of the LHC. The test beam services of these two accelerators, however, will have lower priority starting April 2008 at which time the commissioning of LHC is expected to begin, and thus particularly during the initial LHC commissioning phase uncertainties on beam availability must be expected.
  - This leaves Fermilab as the primary facility to provide necessary variety of particle species in sufficiently wide range of momentum during the next 2 – 3 year period.
- Additional requirements to accommodate the ILC detector R&D beam test activities in the next 3 – 5 year period were identified at the ILC Detector Test Beam workshop in January and through various meetings and workshops throughout the past few months. We thus recommend that the implementation of the following items be seriously considered and implemented on a time scale commensurate with the testing of ILC detectors:
  - An ILC-like bunch train beam time structure of 1ms beam followed by a 200ms blank.
  - A momentum tagged neutral hadron facility for calorimeter beam tests, in particular for the PFA performance and Monte Carlo validations, if meaningful data samples can be collected, taking beam purities and realistic trigger and DAQ limitations into account
  - A common beam test infrastructure for tracking and vertex detectors to provide the same intitial test conditions for all detector technologies. Such a facility should include a large bore, high-field (3 – 6T) magnet, a very high-field (6T) small bore magnet, and a tracking telescope with a pointing resolution of 1 micon.
  - Sufficiently high test beam facility duty factors, which allows for a minimum data collection rate of 1000 particles per minute
  - A common DAQ hardware and software system that can easily integrate additional detector systems
- It should be emphasized that further requirements that could require much more significant facility resources are anticipated in the period after the detector TDRs have been finalized to support the selected ILC detector design and to carry out full-scale system tests.

- Continued, strong support for the test beam activities of ILC detector R&D groups throughout the next 3 – 5 years is critical in making informed decision in ILC detector design in a timely fashion.
- Significant investment in beam test facilities are necessary to accommodate the upcoming ILC detector R&D needs in the next 3 – 5 years and to prepare for the ILC detector prototype calibration and tests after detector TDR submissions.

# 1. Introduction
**Physics Motivations**

The detectors at the International Linear Collider (ILC) are envisioned to be precision instruments that can measure Standard Model physics processes at the electroweak energy scale and discover new physics processes beyond it. To take full advantage of the physics potential of the ILC, the performance of the detector components comprising the experiment must be optimized, sometimes in ways not explored by the previous generation of collider detectors. In particular, the design of the calorimeter system, consisting of both electromagnetic and hadronic components, calls for a new approach to achieve the precision required by the physics. As a precision instrument, the calorimeter will be used to measure jets from decays of vector bosons and heavy particles, such as top, Higgs, etc. For example, at the ILC it will be essential to identify the presence of a Z or W vector boson by its hadronic decay mode into two jets (see for example Ref. [2]). This suggests a di-jet mass resolution of ~3 GeV or, equivalently, a jet energy resolution the level $\sigma/E \sim 30\%/\sqrt{E}$. None of the existing collider detectors has been able to achieve this level of precision.

Many studies indicate that a possible solution to obtain the targeted jet energy resolution of $\sim 30\%/\sqrt{E}$ is the Particle-Flow Algorithms (PFAs) [3]. PFAs use tracking detectors to reconstruct charged particle momenta (~60% of jet energy), electromagnetic calorimetry to measure photon energies (~25% of jet energy), and both electromagnetic and hadronic calorimeters to measure the energy of neutral hadrons (~15% of jet energy). To fully exploit PFAs, the calorimeters must be highly granular, both in transverse and longitudinal directions to allow for the separation of the energy deposits from charged hadrons, neutral hadrons, and photons in three spatial dimensions.

Since PFA requires precision vertex and tracking systems that work coherently with the calorimeter and muon systems, it is critical to not only optimize the calorimeter designs but also to optimize the integrated detector systems to accomplish the physics goals of the ILC.

The ILC physics program requires precise beam instrumentation (BI) for measurements of i) luminosity and the luminosity spectrum, ii) beam energy and beam energy spread, iii) beam polarization, and iv) electron id at small polar angles. The Machine-Detector Interface (MDI) is a key subsystem of the ILC Accelerator and the ILC Detectors. In addition to engineering for the Interaction Region (IR) and the IR magnets, MDI includes the important area of collimation and backgrounds. Many of these BI and MDI topics require test beams in the next 3 – 5 years to complete R&D and to fully develop the engineering design for these systems. It's critical for the ILC Detector community to work closely with ILC accelerator physicists to assure that the MDI and BI are optimized to achieve the ILC physics goals. MDI and BI tests utilize both primary beams at accelerator test facilities and secondary beams at detector test facilities.

**Time Scale Considered in This Document**

The ILC Reference Design Report (RDR) [4] has been released in Feb. 2007. The accelerator project now transitions from a phase dominated by design and R&D to one focused on developing detailed engineering. An ILC Engineering Design Report (EDR) is planned to be submitted in early 2010, to be followed by a 2-year period of negotiations to secure funding approval for a construction start of the ILC and its Detectors by the end of 2012. Seven years of construction is anticipated and the ILC physics program would begin in 2019. ILC R&D will continue, though, and notably there is a very large investment in accelerator test facilities during the next 5 years throughout the engineering and project approval phase. Prototypes will be fully developed, with tests carried out to demonstrate performance requirements and to help refine cost estimates.

ILC detector groups are encouraged to submit Letters of Intent by fall 2008, and current expectations based on the ILC detector roadmap [5] presented at LCWS07 in DESY in early June 2007 are that two of these will be selected to form collaborations and develop full proposals with detailed detector Technical Design Reports to be submitted by the end of 2010 synchronous to ILC accelerator EDR. Similar to the ILC accelerator community, the ILC detectors require significant investment in R&D and test beam activities in the next 5 years to meet this proposed time scale. In fact detector technology choices for vertex, tracking and calorimetry are not as advanced as ILC accelerator systems where baseline choices have already been made.

Fortunately, some detector system choices can be delayed with respect to accelerator system choices, and full advantage of continuing R&D can then be realized before finalizing detector system technologies. Detector system technology choices should be complete by 2012, however, so that the two anticipated ILC detector collaborations are ready to begin construction with a goal to be ready for physics in 2019. Significant increases in ILC detector funding and support for test beam activities are necessary to realize completion of detector TDRs in 2012.

This road map document provides the requirements for each detector subsystem, the current activities and the plans for beam tests through the year 2010 – 2012, at which time detector technologies choices will have to be made.

**2. Facility Capabilities and Plans**

Currently seven laboratories in the world provide eight beam test facilities; CERN PS, CERN SPS, DESY, Fermilab MTBF, Frascati, IHEP Protvino, LBNL and SLAC. In addition, three laboratories are planning to provide beam test facilities in the near future; IHEP Beijing starting in 2008, J-PARC in 2009 and KEK-Fuji available in fall 2007. Of these facilities, DESY, Frascati, IHEP Beijing, KEK-Fuji and LBNL facilities provide low energy electrons (<10GeV). The SLAC End Station-A facility provides a medium energy electron beam, but the availability beyond 2008 is uncertain at this point. IHEP Protvino provides a variety of beam particles in the 1–45GeV energy range, but, due to funding availability, the facility provides beams in only two periods of one month each per year.

The CERN PS and SPS facilities can provide a variety of beam particle species in energy ranges of 1–15GeV and 10–400GeV, respectively. However, given the anticipated LHC turn on, the availability of these facilities beyond November 2007 depends heavily on the LHC commissioning progress. Finally, the Fermilab Meson Test Beam Facility (MTBF) can provide most particles in energy range of 1–66GeV, thanks to the recent beam line upgrade, and protons to 120GeV. This facility is available year-round during the period up to 2011 and probably beyond. Table 1 below summarizes the capabilities of these facilities and their currently known availabilities and plans.

| Facility | Primary beam energy (GeV) | Particle types | Beam lines | Beam Instr. | Availability and plans |
|---|---|---|---|---|---|
| CERN PS | 1–15 | e, h, μ | 4 | Cerenkov, TOF, MWPC | Available, but reduced services during LHC commissioning |
| CERN SPS | 10–400 | e, h, μ | 4 | Cherenkov, TOF, MWPC | Available, but reduced services during LHC commissioning |
| DESY | 1–6 | e | 3 | Pixels | Available over 3 mo/yr |
| FNAL-MTBF | 1–120 | p, e, h, μ | 1 | Cherenkov, TOF, MWPC, Si strips, pixels | Continuous at 5% duty factor, except for summer shutdowns |
| Frascati | 0.25–0.75 | e | 1 | | Available 6 mo/yr |
| IHEP-Beijing | 1.1–1.5<br>0.4–1.2 (secondary) | e<br>e, π, p | 3 | Cherenkov, TOF, MWPC | Available in March 2008 or later |
| IHEP-Protvino | 1–45 | e, h, μ | 4 | Cherenkov, TOF, MWPC | Two one-month periods per year |
| KEK-Fuji | 0.5–3.4 | e | 1 | | Available in fall 2007, for 8 mo/yr, as long as KEKB operates |
| LBNL | 1.5; <0.06; <0.03 | e; p; n | 1 | Pixels | Continuous |
| SLAC | 28.5<br>1–20 (secondary) | e<br>e, π, p | 1 | | Shutdown in 2008-2009, with uncertain plans beyond |

**Tab le 1 Summary of test beam facilities along with their beam instrumentation, availability and plans**

## 2.1 CERN

There are presently four beam lines at two machines; four in the east area of the PS and four in the north area at the SPS. A variety of targets are possible for the PS beams, including one that enhances electron yield by a factor 5–10, but T9/T10/T11 share the same target. For the SPS beams, H2/H4 and H6/H8 share targets. Up to three user areas are possible per beam, although some areas have been permanently occupied by major LHC users. H4 can be set up to produce a very pure electron beam, with energies up to 300 GeV. Low energy tertiary beams are possible in H2 and H8. A summary of the characteristics of the PS and SPS test beams is provided in Table 2. The schematic layouts for the two facilities can be found in [6].

In addition to test beams, there are two irradiation facilities at CERN. The Gamma Irradiation Facility (GIF), based on a $^{137}$Cs source in the former SPS west area, provides 662keV photons at up to 720GBq. While 2007 may be the last year of operation, a new facility is under discussion. A proton and neutron irradiation facility in the PS east hall uses the 24GeV primary protons from the PS to provide a $2\times 2$ cm$^2$ beam spot with $2.5\times 10^{11}$ protons/spill. Neutrons with a spectrum similar to the LHC can be obtained from a beam dump.

In 2007, the PS and SPS test beams will support requests from 47 groups, representing about 1500 users. The PS program will run for 28 weeks, with beam time being devoted to LHC and LHC upgrades (43%), as well as external users (12%). The SPS will operate test beams for 23.5 weeks, supporting LHC and LHC upgrades (52%) and external users (35%). With the start of the high-priority LHC program in 2008, there is considerable uncertainty about the future test beam running schedule. A report from the High Intensity Protons Working Group [7] envisioned three interleaved operational modes for the PS and SPS in the LHC era, including LHC injection, LHC setup (with test beams in parallel), and delivery to other programs, e.g., the neutrino program, and fixed target and test beam experiments. The study suggested about a 50% fraction in the delivery mode in 2008, rising to perhaps 85% by 2011, depending on experience. It remains to be seen what the actually availability will be in the coming years. However, operation of the SPS in test beam mode, and therefore the PS as well, is required to serve several fixed target experiments that are part of the core CERN physics program.

|  | **PS Beamlines** | **SPS Beamlines** |
|---|---|---|
| Momentum range | 1–3.6 GeV [T11]<br>1–7 GeV [T10]<br>1–10 GeV [T7]<br>1–15 GeV [T9] | 10–400 GeV [H2]<br>10–400 GeV [H4]<br>10–400 GeV [H8]<br>10–205 GeV [H6] |
| Spill duration | 400 ms | 4.8–9.8 s |
| Duty cycle | 2 spills/16.8 s | 1 spill/14–40 s |
| Particle types | e, μ, hadrons | e, μ, hadrons |
| Intensity | $1-2\times10^6$ /spill, typically $10^3-10^4$ | $1\times10^8$ /spill |
| Beamline Instrumentation | Beam position monitors, threshold Cerenkov counters |  |

**Tab le 2 Summary of** *available beam lines at the CERN PS and SPS*

**2.2 DESY**

Three test beam lines are available, based on bremsstrahlung photons generated by a carbon fiber in the circulating beam in the DESY II synchrotron. Photons are converted in an external copper or aluminum target, spread into a horizontal fan by a dipole magnet, and then collimated. There are no external beam diagnostics or instrumentation available. However, the T24 area is being dedicated to EUDET, which will provide significant infrastructure. The facility will be down for the first half of 2008, but is otherwise available on a continuous basis. A summary of beam characteristics is summarized in table 3.

**Tab le 3 Summary of available beam lines at DESY**

| **Beam line characteristics** | |
|---|---|
| Momentum range | 1–6 GeV [T21, T22, T24] |
| Particle types | Electrons |
| Bunch spacing | 320 ms |
| Bunch length | 30 ps |
| Rates | 160–1000 Hz |
| Instrumentation | Only EUDET infrastructure in T24 beam line |

### 2.3 Fermilab

The Meson Test Beam Facility (MTBF) has recently completed a major upgrade in anticipation of the needs of the ILC community. By moving the target to shorten the decay path from about 1300 to 450 ft, reducing material in the beam line from 17.8 to 3.4% $X_0$, and increasing the aperture and the momentum acceptance from .75 to 2%, the overall rate has been substantially improved in the new design and the momentum range has been extended below 4 GeV. In addition, the fraction of electrons in the beam has been enhanced. Table 4 summarizes the estimated rates as a function of energy.

**Tab le 4 Summary of low energy beam production rates for Fermilab's MTBF**

| Energy [GeV] | Estimated rate |
|---|---|
| 1 | 1500 |
| 2 | 50K |
| 4 | 200K |
| 8 | 1.5M |
| 16 | 4M |

The Switchyard 120 (SY120) delivers main injector beams to the Meson Detector Building. It must run in conjunction with proton delivery to the pbar source and the neutrino programs. For the purposes of program planning, the MTBF is administratively limited to no more than a 5% impact on these other programs. The Accelerator Division has implemented both 1 second and 4 second spills. Possible configurations are one 4-second spill very minute, 12 hours/day; two 1-second spills every minute, 12 hours/day; and one 4-second spill every two minutes, 24 hours/day. It may also be possible to simulate the ILC beam structure of 1 ms beam followed by 199 ms gap, although with no

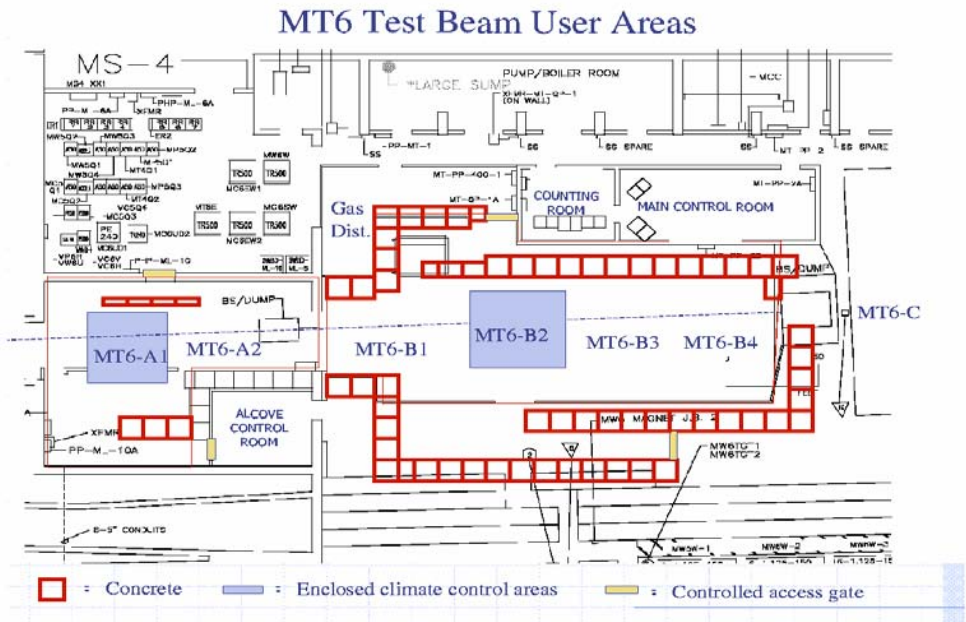

Figure 1 Layout of the user areas of Fermilab's MTBF

micro – substructure to the 1ms beam delivery period [8].

The MTBF test beam area, shown in Fig. 1, is divided into two beam enclosures, although these cannot be operated independently. These enclosures are divided into six user stations and are supported by installed cables, gas lines, offices, and two climate controlled huts. Experiments are also supported by a tracking station, a new TOF system and differential Cherenkov detector, motion tables and video system, and a laser alignment system.

Further enhancements to the Fermilab test bean capability are under consideration. The MCenter beam line, which houses the MIPP experiment, is currently not scheduled. The beam line has very attractive characteristics. Six beam species ($\pi^{\pm}, K^{\pm}, p, \overline{p}$) are available from 1–85 GeV, with excellent particle identification capabilities. The MIPP experimental setup could allow for a better understanding of hadron-nucleus interactions, thereby benefiting our understanding of hadronic shower development.

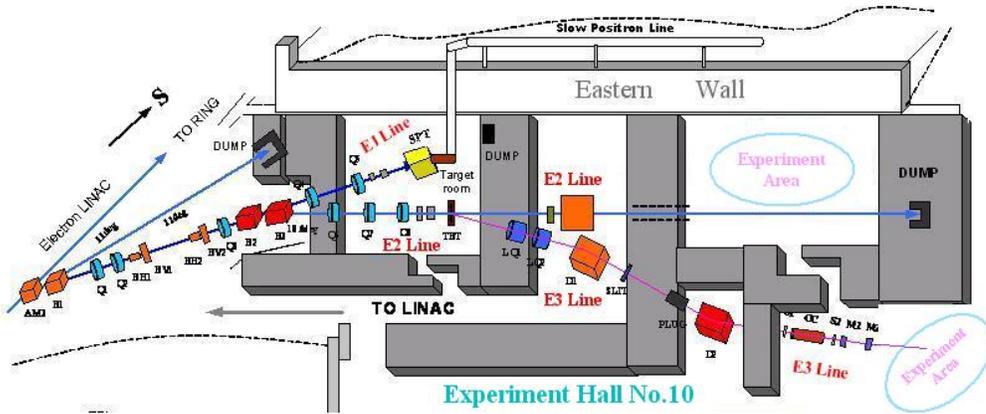

**Figure 2 Schematic layout of the BES test beam area**

### 2.4 BES

Three test beam lines are available at BES, as shown in Fig. 2: two are deliver primary electrons or positrons at 25 Hz to the E1 and E2 experimental areas, while secondary beams at 1.5 Hz are available in E3. The facility is already booked for all of 2007. It will undergo significant upgrade through March 2008, at which point the facility will be available on a continuous basis. Table 5 summarizes BES beam parameters and instrumentations

**Tab le 5 Summary of BES Test Facility Beam Parameters and Instrumentation**

| Momentum range | 1.1–1.5 GeV [E1, E2] |
|---|---|
|  | 0.4–1.2 GeV [E3] |
| Particle types | electrons/positrons [E1,E2] |
|  | electrons, pions, protons [E3] |
| Bunch spacing | 40 ms |
| Bunch length | 1200 ps |
| Rates | 160–1000 Hz |
| Instrumentation | TOF and threshold Cherenkov systems; MWPC with 50% dE/dx resolution |

## 2.5 IHEP-Protvino

At least four high intensity and low intensity beam lines are available at IHEP-Protvino. Beam lines in the BV hall are produced from internal targets in proton synchrotron and have limited intensity. The extracted proton beam is also used to produce high-intensity primary and secondary test beams in the experimental gallery. Test beams are available in two period (April and November-December) for a total of about 2 months/year. Table 6 summarizes the beam line parameters and instrumentation for IHEP-Protvino beam test facility.

Tab le 6 Summary of IHEP Protvino Test Facility Beam Parameters and Instrumentation

| IHEP Protvino beam parameters | |
|---|---|
| Momentum range | 33–55 GeV [N2B] <br> 20–40 GeV [N4V] <br> < 4 GeV [Soft Hadron] <br> 1–70 GeV [N22] |
| Particle types | Electrons, muons, hadrons |
| Bunch spacing | 160 ns |
| Bunch length | 40 ns |
| Rates | 160–1000 Hz |
| Cycle time | 10 s |
| Spill time | 1.8 s |
| Intensity | $10^{13}$ p/cycle |
| Instrumentation | TOF and threshold and differential Cherenkov systems; MWPCs, scintillator hodoscopes |

## 2.6 KEK and J-PARC

There are currently no test beam facilities at KEK. However, the Fuji test beam line is being implemented for fall 2007. This is based on bremsstrahlung photons from 8 GeV high-energy beam particle collisions with residual gas in the KEKB Fuji straight section vacuum chamber. Photons are converted in a tungsten target and the conversion electrons are extracted to an experimental area outside the KEKB tunnel. The expected particle rate is a continuously more than 100 electrons/s over a momentum range from 0.5 to 3.4 GeV. The facility will operate parasitically to KEKB, with availability about 240 days/year.

Plans are developing for test beam facilities at J-PARC, which would be realized no earlier than 2009. Secondary beams with pion and kaon energies between 0.5 and 3 GeV and rates up to $10^5$–$10^6$/pulse may be possible. Table 7 summarizes properties of the Fuji

Tab le 7 Summary of properties of the Fuji test beam facility at KEK

| KEKB test beam parameters | |
|---|---|
| Momentum range | 0.5–3.4 GeV |
| Particle types | Electrons |
| Bunch spacing | 7.8 ns |
| Momentum resolution | 0.4% |
| Rates | > 100 Hz |
| Instrumentation | None |

test beam facility at KEK.

## 2.7 LBNL

Two test beam opportunities are offered, as well as dedicated beam lines for proton and neutron irradiation from the 88-inch cyclotron. A 1.5 GeV electron beam with tunable flux is available at 1 Hz from the injection booster for the ALS. This test area is equipped with a 4-plane beam telescope based on thinned CMOS pixel sensors. In addition, LOASIS is able to supply electron beams via TW laser wakefield acceleration. At present, it is possible to tune beam energies from 50 MeV to 1 GeV. There are also plans to extend the beam line for decreased intensity and to allow testing at different incident angles.

## 2.8 SLAC

A single beam line brings primary electrons from the main linac to End Station A (ESA), with energies up to 28.5 GeV and fluxes varying from $1.0 \times 10^6$ to $3.5 \times 10^{10}$ /pulse. A secondary beam can be produced by putting the primary beam on a Be target in the beam-switchyard and accepting hadrons into the A-line, which makes a 0.5 degree angle with respect to the linac. With 30 GeV primary electrons and the A-line set to 13 GeV, the yield of hadrons is 50% pions, 50% positrons, 0.4% protons, and <1% kaons. Secondary electron or positron beams can also be created using collimators at the end of the linac, with fluxes adjustable down to one particle per pulse. The End Station A facility is well equipped with a shielded area for work with primary beam, and an open experimental region beyond for secondary beams. The beams are well instrumented, with unique characteristics as summarized in Table 8.

The present End Station A program operates parasitically to PEP-II operations. Anticipating the end of the B Factory running in September 2008, the user-based test beam program in End Station A (ESA) will complete in Summer 2007, though some ILC tests will continue in ESA until September 2008. In 2009, the downstream 1/3$^{rd}$ of the linac will be used for the LCLS project with no plans for delivering test beams to ESA. The South Arc Beam Experiment Region (SABER) has been proposed as a follow-up to the Final Focus Test Beam Facility (FFTB). SABER would use the first 2/3 of the linac to deliver compressed, focused, primary electrons and positrons at 28.5 GeV to the south arc experimental region. This space may be suitable for smaller scale R&D experiments. SLAC is also considering an extension of the SABER proposal that would provide 28.5 GeV primary beams to the A-line, thereby restoring capability for both primary and secondary beams into End Station A. SABER is scheduled for operation in 2010, so a user test area could be restored in either the south arc or End Station A shortly afterwards.

Tab le 8. Summary of available beam lines at the SLAC End Station A

| End Station A beam line | |
|---|---|
| Momentum range | 28.5 GeV<br>1–20 GeV secondaries |
| Particle types | Primary electrons<br>Secondary electrons/positrons, pions |
| Bunch spacing | 100 ms |
| Rms Bunch length | 300–1000 μm |
| Rms Spot size | $(x, y) = (100, 600)$ μm primary beam<br>(x,y) = (3mm, 3mm) secondary beam |
| Energy spread | 0.2% |
| Rates | $1.0 \times 10^6$ to $3.5 \times 10^{10}$ primary electrons/pulse at 10 Hz<br>1–60 secondary electrons or positrons/pulse at 10 Hz<br>1–10 secondary hadrons/pulse at 10 Hz |
| Instrumentation | TOF, threshold Cherenkov systems |

## 3. Detector R&D Group Needs and Schedule

This section describes the needs and schedule of detector R&D groups. The section contains short summary of activities, upcoming plans of each R&D group within the subdetector group. This section also contains short summary of specific needs as a function of time scale along with the justification for the needs in the followings:

- Beam: particle species, momentum, spatial spread, momentum bites, beam time structure, polarity selection
- Beam line instruments: necessary instrumentation, precisions of the beam line instruments, software
- Equipment: magnets – field strength, bore size and additional specific needs (such as split coil), cranes (capacity, etc), movable table (sizes and functionalities, motion ranges, etc)
- Short summary of expected time needs

### a. Machine-Detector Interface and IP Beam Instrumentation

Luminosity and energy reach are the key parameters for the ILC. The ILC's capability for precision measurements, though, distinguishes it from the LHC. The ILC physics program requires precise IP Beam Instrumentation (IPBI) for measurements of i) luminosity and the luminosity spectrum, ii) beam energy and beam energy spread, iii) beam polarization, and iv) electron id at small polar angles [9 – 11]. The Machine-Detector Interface (MDI) is a key subsystem of the ILC Accelerator and the ILC Detectors. In addition to engineering for the Interaction Region and the IR magnets, MDI includes the important area of collimation and backgrounds. Many of these IPBI and MDI topics require test beams in the next 3-5 years to complete R&D and to fully develop the engineering design for these systems. Later, beam tests will evolve from testing prototype systems to test and validate final systems for the ILC. Desirable test beam parameters and facility capabilities for MDI and IPBI beam tests are summarized in Table 9. In general, it is desirable to have high energy, low emittance beams with similar bunch charge, bunch length and time structure as ILC beams. Adequate space, beamline instrumentation and crane availability are also desirable. These desired features cannot all be realized in a single facility. Some MDI and IPBI tests also require specialized capabilities, including the availability of low intensity secondary beams.

**Table 9. Summary of Desired Features for MDI/IPBI Test Beam.**

| Parameter | Test Beam Requirement |
|---|---|
| **Energy** | 1-30 GeV |
| **Charge/bunch** | $(0.2 – 2.0) \cdot 10^{10}$ |
| **Repetition Rate (Hz), Pulse Structure** | 5-10Hz; ILC train with 150-300ns bunch spacing |
| **rms Pulse Length** | (100-1000) μm |
| $\gamma\varepsilon_x, \gamma e_y$ **(mm-mrad)** | as low as possible; $\gamma\varepsilon_x, \gamma e_y$ <(300, 20) acceptable for most tests |
| **rms (x,y) Spotsize** | <1mm; (5-20) μm for some tests |
| **rms Energy Spread** | <(0.5-1)% |
| **Momentum precision** | (100-1000) ppm for some tests |
| **x,y,z space required** | 0.5m, 0.5m, (0.5-30)m |
| **Instrumentation needs** | Q, x, y, x', y', $\sigma_x$, $\sigma_y$, $\sigma_z$, E, $\sigma_E$ |
| **Crane capacity** | Up to 15 tons |

At ILC, the luminosity will be precisely measured using Bhabha events. A precise and compact calorimeter, LUMICAL, in the very forward direction at polar angles of 40-120mrad will be used for this. At smaller polar angles of 5-40 mrad, excellent electron identification is required in the BEAMCAL detector to veto copious 2-photon events that present a serious background to SUSY searches. BEAMCAL resides in a difficult radiation environment and is hit by a large flux of low energy $e^+e^-$ pairs, depositing 10's of TeV per bunch crossing in the detector that can be used for fast luminosity tuning.

The determination of particle masses, e.g. for the Higgs boson or top-quark, requires a precise determination of the beam energy to 100 parts-per-million (ppm). Beam energy spectrometers utilizing precise beam position monitors (BPMs) and synchrotron stripe detectors are being designed to achieve this. The synchrotron stripe detectors also have capability to measure the beam energy spread. Very precise parity-violating asymmetry measurements at the ILC allow sensitive probes for new physics, and require polarimetry measurements to better than 0.25%. Compton polarimeters are being designed to achieve this.

The GamCal measures the flux of beam-strahlung gammas at z = 190m from the IP. A thin $10^{-4}$ radiation length foil converts a small fraction of the gammas to electron positron pairs, which are then swept into the GamCal detector by a dipole magnet with PT=0.25GeV. The GamCal is needed for optimization of the instantaneous luminosity. This detector requires a high energy electron test beam (E>20 GeV) to compare the measured backgrounds from the electron beam with our simulations during CY08-09.

There is a large community actively engaged in R&D for MDI and IP Beam Instrumentation. This includes both detailed design and test beam activities, which are vital to validate the designs. An overview of IPBI and MDI test beam activities follows.

The FCAL collaboration [12] is leading the effort on very forward region calorimetry for the LUMICAL and BEAMCAL detectors. Recent test beam activities have focused on radiation damage studies for potential BEAMCAL sensors, such as silicon or diamond, at the DALINAC accelerator at the TU Darmstadt. DALINAC provides a 10 MeV electron beam for these studies, similar to the average energy of shower electrons in BEAMCAL. These radiation damage and sensor studies are continuing. Future BEAMCAL test beam studies should also include measurements of electron id efficiency and shower fluctuations for high energy multi-GeV electrons.

At SLAC's End Station A, a program [13] to test components for the Beam Delivery System and Interaction Region is underway. The ESA beam energy is 28.5 GeV and has similar bunch charge, bunch length, and bunch energy spread as planned for ILC. A major component of the program there is developing protoype BPM and Synchrotron

Stripe energy spectrometers [14] in a common 4-magnet chicane. The ESA program should continue through FY08. Beyond FY08, it is uncertain whether the ESA facility will be available. Another test facility at SLAC, SABER, should become available after 2008 with similar beam parameter capabilities as currently available in ESA. SABER could be useful for a number of ILC beam tests, but accommodating energy spectrometer prototypes will be difficult due to the limited space and infrastructure available.

A wide research program for beam instrumentation is carried out at the ATF [15] at KEK. The ATF delivers an 1.3-GeV electron beam with micron-sized bunches. Three bunches with 150 ns or two bunches with 300 ns between bunches can be delivered as a train. High resolution nano-BPMs are being tested there that will be important for the BPM energy spectrometer. Laserwire diagnostics [16] are also being tested and have potential use for beam energy spread measurements. The ATF2 project [17] will extend the extraction beamline of ATF to a final focus beamline prototype for the ILC. The goal is to achieve a beam size of 35 nm and nanometer stability simultaneously. Development of nano-BPMs and laserwires will continue at ATF2 in 2008 and beyond. Laserwires are also in use at PETRA and will be further developed there for PETRA III after 2008.

Other IPBI and MDI topics have active R&D programs requiring (or may be requiring) test beams. These include: electro-magnetic interference [18], collimation and backgrounds (ex. collimator wakefield program at SLAC [19] and plans for material damage studies [20]), FONT IP BPM backgrounds [21], polarimetry (ex. DESY polarimetry R&D [22]), and the stability of the magnetic center of IR quadrupoles.

3.2 **Vertex**

Currently, vertex detector prototypes are being tested in laboratory facilities and test beams round the world, mostly under conditions far from those to be encountered at ILC. This phase will continue until R&D collaborations start to build prototype detector sub-assemblies (a significant fraction of a complete detector of their preferred design, maybe 10% of the full assembly of ~1 Gpixel, including realistic mechanical mounting, electrical and cooling services).

Once this stage is reached, starting maybe in 2010 and continuing somewhat beyond 2012 for candidate technologies for use as first-generation detectors at ILC, there is general agreement in the vertex detector community that the performance of these sub-assemblies should be evaluated in a *common infrastructure*, for two main reasons. Firstly, the required facilities will be expensive and technically non-trivial, making duplication undesirable in terms of cost effectiveness. Secondly, a facility that is used by everyone will lend itself to objective testing and comparison of results. Without this, there would be a risk that incorrect choices of detectors could be made, due to misleading differences in the evaluation and presentation of results from different test systems.

The vertex community, subject to final approval by the WWS-OC, is in the process of setting up a coordination group, analogous to the task forces of the accelerator R&D, in order to plan and develop the common infrastructure. Without a coordinated approach, there is a risk that requirements of one group may be overlooked in the pressure to move forward by another, despite the claim by that group to be developing a shared 'user facility'. The risks are exacerbated by the wide range of technologies being considered for vertex detectors at ILC. A test facility that would provide all that is needed for the FPCCD evaluation may be lacking in features essential for testing chronopixels, and *vice versa*.

The shared test beam will form an essential part, but only a part, of the common infrastructure. What is needed is an agreed plan for the overall facilities, some of which will presumably be shared with the ILC tracking R&D groups. The formal planning of these facilities depends on establishing the relevant coordination groups. In the meantime, this report should be read as a provisional indication of what is likely to be needed, based on incomplete discussions in the community. It is too early to recommend specific plans, which would surely overlook the needs of some of the R&D groups.

Starting with the beam itself, while a variety of particle types and momenta are currently used, the definitive testing will require a hadron beam of some tens of GeV, in order that the few micron precision of the detectors can be measured without excessive smearing induced by multiple scattering. Since it is unlikely that a suitable test beam will be available for ILC work at CERN on the required timescale, the most probable option is Fermilab, where it seems that the Meson Test Beam Facility (MTBF) can probably be adapted to do the job. Firstly, in order to test some but not all of the vertex detector candidate technologies, it will be necessary to replicate the coarse time structure of the ILC (spills of 1 ms duration at 200 ms intervals) through the extraction period (~4 s) of the test beam.

Preliminary discussions suggest that this could be done at Fermilab's MTBF for moderate cost. Secondly, it seems that it will be desirable, for tests of 2-hit resolution and to study the integrity of DAQ for groups of closely spaced hits simulating particles in cores of jets, to be able to achieve a focused beam spot as small as 1 mm diameter. Should this be impossible with the required flux (which still needs to be quantified), other options for testing these aspects of the detectors should be explored. The precise requirements during the 200 ms 'beam-off' intervals need careful thought. It may be that integrated flux during these periods of 10% of the 1 ms 'beam-on' flux will be acceptable, which would simplify the design of the extraction system. These and other questions will remain open till the coordination work gets started.

Different detector technologies will have different requirements for a beam telescope. Logically, just one such telescope should be constructed, to satisfy the most severe of those requirements. These will probably be determined by the chronopixel option, and others in that family. Due to their sensitivity for periods of only ~50 ns at 300 μs intervals during the spill, it will be necessary to tag the subset of particles which arrive during their sensitive time. An alternative option in principle would be to replicate the ILC beam fine structure (3000 instantaneous bunches during the 1 ms spill) but this is unrealistic in a test beam, for a number of reasons. A more realistic approach will be to use a high precision beam telescope (presumably about 6 planes of silicon microstrip detectors) that provides ~10 ns timing precision on each incident particle. Such a telescope will probably become available from previously operating test beam facilities or fixed target experiments, on the required timescale.

There is general agreement that the detector sub-assemblies should be tested in an appropriate (3 – 6 T) magnetic field. There are many issues related to the detector operation which could cause nasty surprises if an ILC vertex detector were to be built without having been subjected to such tests. From the initial charge collection (which may be significantly affected by Lorentz angle effects in the ILC solenoid field) to mechanical forces induced by the power pulsing at a frequency of 5 Hz, there are numerous effects which could cause severe problems. Ideally, the sub-assembly will be mounted in the ILC orientation in a rotatable solenoid of the split pair type, so that the beam can traverse the detector cleanly in the direction normal to the cylindrical elements, as well as over a range of oblique angles, passing through the open ends of the magnet coil.

The required dimensions of such a magnet will need to be established by the coordination group, and it may be that a suitable one can be found. If not (and the important split coil feature may make this difficult) the commercial production of such a magnet would be entirely affordable. It is also possible that this magnet may not be needed, if the vertex detector prototypes can be mounted inside the much larger solenoid that will be required for the tracking detector large scale prototypes. However, this could create scheduling and other problems, which will emerge from discussions between the coordination groups. If it turns out that two solenoids are needed, the relatively small size of the vertexing coil means that the overall cost will not be much greater than for the tracking solenoid alone.

Having each type of vertex detector passing through the same evaluation infrastructure will provide an important opportunity for performing studies which, if done by different people in different laboratories, would be difficult to compare. Firstly, there have frequently been discrepancies between the estimates of material budget (detector thickness in radiation lengths) and what was actually built. The true values can best be determined by weighing and by measuring attenuation of an X-ray beam in which the detector is rotated to map out the full solid angle.

Secondly, tolerance of background hits can be measured and compared by introducing a standard strength gamma source into the solenoid enclosure while taking beam data. Thirdly, mechanical effects of pulsed power will become visible (indirectly) from the quality of the fitted tracks, and (directly but only with partial coverage)) by optical monitoring of visible parts of the nested detector system. Given that the sub-assemblies will presumably consist of partial barrels and partial end-disks, it may be that optical monitoring of all layers will be possible. Fourthly, radiation hardness can be determined by repeatedly testing the sub-assembly after exposure to increasing doses of both ionizing radiation and neutrons, made in the same irradiation facilities under identical conditions for each candidate detector.

In addition to studies in the high field solenoid with the test beam, the shared infrastructure should include a calibrated facility (anechoic chamber, RF sources) for systematic studies of the EMI sensitivity of each detector sub-assembly. From studies currently under way in test beams (eg the SLAC ESA beam) a picture is emerging of likely levels of EMI that will escape into the detector environment at ILC, mostly due to such effects as apertures needed to transport signals from BPMs near to the IP. Putting together this information with the results of the EMI sensitivity measurements, it will be possible to determine which technologies could be in trouble under ILC operating conditions.

The facilities in this shared infrastructure should be used to provide a comprehensive evaluation of each detector technology, so that by about 2012, when first generation vertex detectors will need to be selected for use at ILC, the community will have reliable data on which to base the decisions. Beyond 2012, it is most likely that these facilities will continue to be needed, since superior pixel technologies will surely be emerging, that were too adventurous for ILC startup, but which will be serious candidates for upgrade detectors. The combination of accessibility of the inner region, which is embodied in all the detector concepts, and the down periods enforced by the push-pull operation, will provide the detector collaborations with convenient opportunities to make such upgrades.

Judging from past history, one might expect superior new technologies to take over as next generation ILC vertex detectors, at intervals of around 3 years in both of the detector systems. It is therefore important to make careful long term plans for the support and maintenance of the vertex detector common testing infrastructure, of which the test beam will comprise the most essential element. If located at Fermilab, this and the companion tracking test facilities could provide some of their most important and enduring contributions to the ILC experimental program.

3.3 **Tracking**

The WWS R&D panel completed a review of the tracking detector R&D activities at the ACFA ILC meeting in Beijing in February 2007. Four tracking groups presented their needs for beam tests during that review. Two groups, the Linear Collider TPC (LCTPC) group and the Cluster Counting (CluCou) drift chamber group are pursuing gaseous based tracking detectors. The Silicon for the Linear Collider (SiLC) collaboration and the Silicon Detector (SiD) concept pursue silicon based tracking.

To date the TPC collaboration has carried out substantial R&D in the DESY and KEK test beam lines with small prototypes (SP). The collaboration is now moving towards building large prototypes (LP). Plans are well underway for testing the first large prototype (LP1) at DESY. The LP will be about 1m in diameter. These beam tests will be incorporated in the EUDET R&D plans over the next two to three years. The intent is to test LP1 in realistic beam conditions. This R&D program will initially run at the EUDET facility located in a 6 GeV electron beam at DESY. For these initial efforts 6 GeV electrons, combined with cosmic ray tests, will be sufficient, but ultimately higher momenta test beams will be needed so that LP1 will have to move to a high energy beam at Fermilab or CERN. LP1 is foreseen to start taking data the latter part of 2007, depending on when the field cage and electronics are ready and on how fast the endplates can be developed. Several designs are considered for the endplates: a GEM solution, Micromegas and a SiTPC solution. The testing and data taking of the GEM and Micromegas would last until the end of 2008, at which time some SiTPC prototypes are expected to be ready.

The collaboration hopes to move towards testing the large prototype 2 (LP2) as soon as possible. Testing of the LP2 would follow LP1 at higher momenta at Fermilab or CERN. In principle part of the infrastructure provided by EUDET would be relocated at Fermilab or CERN. It is clear, however, that the TPC community would benefit from the availability of a large bore, high field solenoid. Two magnets are currently available or will become available rather soon. The AMY magnet has a field of 3T and a cryostat with an inner diameter of 2.2m. The TRIUMF magnet has a field of 2T, an inner diameter of 1m and a length of 223cm. How cost effective it is to move either one of these magnets to the CERN or Fermilab beamline is unclear. Furthermore, it has not been established if the field strength is adequate to carry out all tests.

The CluClou collaboration hopes to establish in detail the viability of this approach to tracking for the ILC in the coming year. They have no predefined requirements on test beams as of yet and anticipate that they will be able to use the existing infrastructure once the need arises.

The SiLC collaboration is exploring what silicon tracking system will be needed in conjunction with a gaseous central tracker, and intends to study the requirements for forward silicon tracking, whether in conjunction with gaseous or silicon central tracking systems. This work will need large scale prototypes in appropriate beam tests, to establish a viable solution to this most important problem. The SiLC collaboration envisions collaborative activity with the European and Asian silicon and TPC communities. The collaboration is currently testing long ladders in the DESY test beam but would like to move to Fermilab for test with higher momentum particles. Due to the nature of their program -- testing ladder assemblies with the large TPC prototype -- their plans are closely linked with the TPC program. Looking forward to the time period 2010-2011, they intend to participate in using the common infrastructure that hopefully will have developed by then. They will test large-scale prototypes of their ladders with different lengths of sensors and different electronics options, as well as prototype forward disks as stand-alone and in conjunction with a TPC prototype.

The SiD silicon tracker group is planning to perform beam tests for their prototypes starting in 2008 at Fermilab. A gradual move to larger scale system tests is foreseen by 2010, the latter to include a large disk as well as a barrel prototype. At the moment they do not foresee a need for a large bore, high field magnet. Most sensor studies can be carried out with a small bore, high field magnet.

Initial studies of the four R&D groups' prototypes can be conducted at low energy electron facilities, such as DESY, but these studies do not suffice. There is a clear need by all groups to conduct large scale prototype performance tests. This would require a move to facilities with available high energy tagged particle beams. They will also need a large bore high field (3 – 6 T) magnet that can accommodate large-scale prototypes at the facility to study detector performances and vibration issues coupled with pulsed power. Finally, a beam structure of 1ms beam followed by a 200ms of blank is needed to mimic ILC beam structure as close as possible.

3.4 **Calorimetry**

As a precision instrument, the ILC detector calorimeter will be used to measure jets from decays of vector bosons and heavy particles, such as top, Higgs, etc. It will be essential to identify the presence of a Z or W vector boson by its hadronic decay mode into two jets. This suggests a di-jet mass resolution of ~3 GeV or, equivalently, a jet energy resolution $\sigma/E \sim 30\%/\sqrt{E}$. None of the existing collider detectors has been able to achieve this level of precision and therefore the testing of alternate possibilities for detector solutions in a test beam is of utmost importance.

Given the critical role of the calorimeter system for the overall detector concept, the complexity of its development to meet the stringent requirements for ILC physics, and the need to cover a wide energy range with large statistics, calorimeter groups dominate the need for beam tests. There are a total of five ECAL groups, three digital hadron calorimeter groups and four analog calorimeter groups. All of these calorimeter R&D groups work within the CALICE collaboration, with the exception of Oregon-SLAC-BNL (OSB) Si/W ECAL and the dual readout crystal calorimeter of the forth detector concept.

Since variety of beam particle species in wide range of momenta are needed to understand calorimeters at the required level, the calorimeter R&D effectively determines the specifications of any ILC detector R&D test beam facility. In addition, even prototype detectors for calorimetry can have over 400,000 readout channels, just to cover the volume of a single particle hadronic shower with the envisaged granularity, and put a burden on the infrastructure of any test beam facility. The ILC calorimeter and muon detector community has summarized the requirements for test beams in a Fermilab technical memorandum [23].

The CALICE collaboration completed a six week electron and hadron beam test of its Si-W ECAL, a scintillator-Steel AHCAL and scintillator-steel tail-catcher (TCMT) at CERN's H6 beam line summer 2006. This beam test was preceded by several exposures of the ECAL and AHCAL to 6GeV/c at DESY in 2005 and 2006. One sensitive layer of the TCMT was exposed to 3 GeV/c electron beams at DESY in 2005 and to 16 GeV/c pions and muons as well as 120GeV/c protons at Fermilab's MTBF in 2006 (T957).

This prototype detector configuration of the CALICE collaboration has been granted of two periods of beam for two weeks each at CERN's H6 beam line in summer 2007 for the completion of detector commissioning and high energy hadron beam test runs. The complete set-up is expected to move to Fermilab in late 2007 for a very low to medium energy hadron and electron beam exposure, long-term detector performance and further combined PFA shower shape runs. The group will also explore the possibility of momentum tagged neutral hadron beams at the Meson Center beam line at Fermilab. This system, including its movable mechanical support, is going to be shared with both the DHCAL groups and the Asian scintillator calorimeter groups for their combined performance tests.

The RPC digital calorimeter (DHCAL) group in the US has performed a chamber characteristics experiment at MTBF (FNAL–T955) in Feb. 2006. They tested 3 RPC chambers with 120GeV/c proton beams. The GEM DHCAL group conducted a beam test to characterize their 30cm×30cm GEM chamber in Mar. – Apr. 2007. These two groups plan to perform a "Vertical Slice Test" in summer 2007 to test ANL-FNAL joint developed DCAL digital readout chip [24] and the SLAC developed kPix analog readout chip [25, 26]. This test will be followed by a full scale test using a 40 layer, 1m$^3$ prototype in early 2008, as part of the CALICE collaboration beam test effort. The 1m$^3$ prototype test of DHCAL depends heavily on funding.

The Asian scintillator-W ECAL group completed a prototype test at DESY in February-March 2007. The test module will then be upgraded and tested at Femilab in late 2007 or early 2008, exploiting the CALICE detector stack and scintillator HCAL readout systems, which are expected to be taking data at Fermilab in the same time scale. New hadron calorimeter test modules will also be constructed with scintillator tiles and possibly strips, as part of the CALICE AHCAL effort. They should be equipped with time-resolving electronics for the study of the neutron component of hadron showers, which has not been possible with the existing prototypes. These tests will be performed in 2009 and 2010. Other Asian ECAL groups have joined the CALICE collaboration and are planning to conduct beam test jointly using the existing stacks.

In addition to the traditional purpose of beam tests, the performance test of the prototype and the feasibility of the technology, calorimeter beam tests are needed also to validate Monte Carlo models used to develop the Particle Flow Algorithms (PFAs). The entire calorimeter, consisting of ECAL, HCAL and TCMT, needs to be tested in a wide variety of test beam configurations in order to gauge the performance of the given technology. The requirements span the range of particle types (electrons, pions, muons, and protons), momenta (1 GeV-120 GeV) and several angles of incidence and wide range of rates. Recently a possibility of the momentum tagged neutral hadron beam has been discussed at Fermilab's Meson Center beam line.

In addition to the use of MC models, the test beam data may in principle be used to generate extensive libraries of hadronic showers, which could be used to test some aspects of the predicted ILC detector performance. Large data samples have to be collected, since the models differ mostly in the description of the peripheral tails of the shower, which are however vital for the evaluation of the ultimate particle flow performance. Based on the experience made at CERN in 2006, we estimate that a total running time of about 6 weeks (in 2 or 3 portions, including calibration) is needed to test a given ECAL HCAL configuration. This assumes an effective data taking rate of 60 Hz averaged over the duty cycle of the machine and a beam availability of 60-70%; for different values the estimate scales accordingly.

Finally, a beam time structure that mimics that of the ILC, 1ms beam followed by a 200ms blank, is needed to test the functionality of the pulsed power driven electronics.

3.5 **Muons and Tail Catchers**

The FNAL, Indiana University, Northern Illinois University, Wayne State University and University of Notre Dame jointly constructed 4 planes of strip scintillation counters (each plane consists of 64 strips in an area 1.25m X 2.5m) and completed initial beam tests at Fermilab's MTBF (FNAL–T956) in the Fall of 2006 with 256 strips from 4 planes. The purpose of the beam tests was to determine baseline characteristics of the detectors that employed wave-length shifting (WLS) fibers (Kuraray Y-11) glued to extruded scintillator then thermally fused to clear fiber to exit the expected magnetic field region where the WLS shifted scintillation light was converted to fast pulses using multi-anode photomultiplier (MAPMTs) 64 channel Hamamatsu H7546B tubes. The test results have established that in excess of 6 photo-electrons are detected from the passage of charged mesons, protons and muons through strips 1 cm thick in the beam direction. The MAPMTs performed very well running at a gain of about 2M with HV ~960V. Typical pulses had rise times of 5 – 10 ns and yielded a measured charge of 5 - 10 pC.

In addition to the direct measurements of the scintillator pulses, efficiency and transmission of the WLS and clear fiber assemblies, the group studied the calibration/gain/response of the MAPMTs and the amplifier and digitization process and developed both hardware and software to collect the data, characterize the hardware and analyze the data. While the DAQ hardware was not customized to test beam usage, it was useful for a first measurement of ~ 24 channels. To do more measurements and testing of ILC Muon/Tailcatcher scintillator prototype detectors, electronics, etc. there are obvious areas where we want improvements, simplification and more modern technology.

For sometime we have anticipated testing pixilated solid-state avalanche photo-diodes to detect the WLS scintillation light. In fact before we concluded our Meson Test Beam Facility (MTBF) testing in September 2006 we added several days of effort to test, with high energy charged particles, a few new Si detectors manufactured at the IRST-Trento INFN Laboratory facilities in Italy. Our colleagues, Prof. Giovanni Pauletta from INFN Trieste-Udine and his student, with modest help from us, were able to get interesting preliminary data that looks very promising toward indicating the use of such devices as the photo-detectors for an ILC scintillator-based Muon-detector/Hadron-Tailcatcher. At the present time we are in the planning stages for developing a few more planes 1.25m X 2.5m detectors to test with high energy beam.

The group expects to continue to test strip-scintillator detectors equipped with multi-pixel photon detectors, using Minerva electronics or an alternative, at MTBF in 2007/2008. The group's expected use of the beam will consist of a number of two week periods during 2007/2008. In the second half of 2007, the group expects to test 1 or 2 new planes and in 2008 perhaps one full-scale prototype, 2.7m (h) x 5.7m (w).

In addition, there may be a desire to test RPC chamber prototypes in MTBF in the future after initial tests are done at SLAC and elsewhere. One of the objectives is to measure their beam rate response, which may require high intensity beam. However, details of this beam test are not yet well known. In Asia, it is anticipated that the GLD Muon effort will focus on further tests of MPPCs in Japan and probably with strip scintillator detectors in CERN and DESY in 2007. In 2008 and beyond they may want to use the Fermilab facility. The size or scope of detectors that they would bring to Fermilab for testing is not yet known. In Europe, based on discussions with Marcello Piccolo (INFN/Frascati) and Giovanni Pauletta (INFN Udine/Trieste), there is interest in further testing Italian SiPMs installed in strip-scintillator planes.

Given the fact that these muon detectors are expected to perform with other detectors in the path of muons, these beam tests can occur concurrently with other detector beam tests. The particle species and energies needed in muon plane testing are already mentioned in the other groups' needs.

## 4  Computing, Simulation and Software Needs

The primary task of the test beam efforts is clearly the development of the actual detector hardware. However, these measurements provide also a testing ground to develop software concepts for the processing of real data. The more that with the 'next generation' of prototypes the detector components themselves and the subsequent data acquisition systems move closer to what will be installed in a tentative ILC Detector. The test beam efforts do allow for a first major development step towards the establishment of a full data processing including the handling of conditions data and the application of grid tools for data distribution and data reconstruction.

LCIO is widely used within the ILC community for detector studies. R&D groups have expressed their desire to process their data using a common data format and where possible using a common software framework. This would facilitate at a later stage the performance of combined test beams and to propagate test beam results into full detector studies. The application of the ILC software requires the development of strategies for the handling of raw data including core data needed for the analysis of the test beam measurements by a wider user community and 'service data' which are more useful for the actual DAQ debugging . These strategies are currently under consideration.

The next generation DAQ systems won't deliver the data organized as events but rather will deliver complete bunch trains. It will be the task of a dedicated server program to decompose these bunch trains into events onto which analyses are usually based on. It is a matter of discussion whether this server program should be an integrated part of a (central) DAQ system. The current tendency is that it will be part of the DAQ in order to provide that all analyses can start one a unique set of files leaving intermediate information to experts for debugging purposes. The format of the output file should be compatible with the file format used in all ILC studies.

To establish the data processing and the interplay between DAQ systems and the ILC Software, it is highly desirable that test beam facilities deliver a beam structure as close as possible to the bunch/beam structure foreseen for the ILC. Beyond the requirements to the actual facilities, test beam activities have to be accompanied by more infrastructural tools. These tools are for example the offering of a database service as well as providing resources in the grid for data storage, transfer and processing. The latter services may be provided by laboratories which do not operate a test beam facility.

A successful analysis of the test beam data requires access to beam relevant parameters such as e.g. beam energy, collimator settings etc. These data will have to be fed into the data stream of the actual test beam data. Here it is important that the test beam facilities provide convenient interfaces to these parameters. A detailed demand on needed parameters should be formulated beforehand to the operators of the test beam facilities by the experimental groups.

In particular the measurements with high granular calorimeters will deliver important information about the precision which is expected to be achieved with an ILC detector. The high granularity demand novel approaches to clustering algorithms. In particular for hadronic cascades these algorithms can only be optimized in case the underlying model for hadronic showers is correct. The imaging capabilities of the calorimeters will deliver very detailed pictures of hadronic showers. These days the models available in simulation packages as GEANT3, GEANT4 and Fluka lead to considerable different predictions for typical observables as the shower radius as these models have not yet been optimized for high granular calorimeters

The test beam measurements provide a unique opportunity to verify hadronic models on prototypes which will have similar characteristics as the full detectors later on. The correct description of the data demands a close collaboration between the experimental groups and the developers of these software packages. The development and qualification of appropriate models will require significant computing resources primarily in terms of grid resources. Laboratories all around the world interested in ILC physics are requested to provide these resources together with the corresponding support.

## 5 Summary of Detector R&D Groups' Requests to Facilities and Time Scale

The needs from various detector R&D groups that could possibly require significant resources and/or long lead time can be summarized as follows:

- ***Common beam test infrastructure for tracking and vertex detectors to provide uniform facility that can be used for performance evaluation and informed decisions in technology choices for ILC detector TDRs.***
    - o Detailed specification of such facility and the coordination of the use of the facility need to be finalized soon after the IDAG director is chosen.
- ***Tagged neutral hadron facility for calorimeter beam tests, in particular for the PFA performances and MC validations***
- ***ILC like beam time structure of 1ms followed by 200ms blank***
    - o All sub-detector groups require this feature.
    - o The specification on the level of mimicking is needed. Is the micro-structure within the beam bunches within the 1ms beam period needed or mimicking the macro structure of 1ms beam followed by 200 ms blank sufficient?
- ***Sufficiently high test beam facility duty factor***
- ***Possible common DAQ system***

| Subdetector system | Number of groups | Particle Species | P (GeV/c) | B (T) | Nweeks/ year | ILC Time Structure | Note |
|---|---|---|---|---|---|---|---|
| BI&MDI | 16 | e | =<100 | - | 64 | - | Mostly low E electrons |
| Vertex | 10 | e, π, p, μ | =<100 | 3 – 6 | 40 | Yes | |
| Tracker | 6 TPC+2Si | e, π, p, μ | =<100 | 3 – 6 | 20 | Yes | |
| Calorimeter | 5ECALs 3DHCALs 5 AHCALs | e, n, π, K, p, μ | 1 - >=120 | Some needed | 30 – 60 | Yes | |
| Muon/ TCMT | 3 | e, π, μ | 1 - >=120 | - | 12 | - | |

**Table 10. Summary of currently known ILC detector R&D group test beam needs; It should be noted that most calorimeter R&D groups are in CALICE collaboration.**

Table 10 above summarizes the currently known ILC detector R&D groups and the needs for each subdetector system. The information on this table is only a snapshot of the current beam test activities and plans, thus will change as circumstance changes.

## 6 Conclusions

The International Linear Collider provides an important opportunity for precision measurements that would take our understanding of nature one step further. The design of accelerator is making impressive progress. The ILC detector R&D activities start to advance much more expeditiously to complete ILC detector TDRs synchronously to the accelerator and to complete in 2010 time scale for a construction of detectors in 2010. Thus the detector R&D beam test activities will increase, and the demand to the facilities will grow significantly. We presented in this document the roadmap for ILC test beam in the next 3 – 5 years, precisely the time scale for completion of detector TDRs in an informed manner. We believe this document provides necessary information to the leaders of ILC physics and detector community and the facility managers, as well as funding agencies to prepare for the anticipated large influx of beam test activities both in providing sufficient resources to detector R&D groups to carry out beam tests and to facilities to prepare adequately to accommodate the needs in these activities.